\documentclass[runningheads]{llncs}
\usepackage{graphicx}
\usepackage{algorithm}
\usepackage{algorithmic}
\usepackage[algo2e]{algorithm2e}
\usepackage{graphicx}
\usepackage{graphics}
\usepackage{subfig}
\usepackage[colorlinks]{hyperref}
\usepackage{diagbox}

\usepackage{float}
\usepackage{environ}
\NewEnviron{equations}{%
\begin{equation}
\scalebox{1.5}{$\BODY$}
\end{equation}
}
\usepackage[font=scriptsize]{caption}
\newcommand{\repeatthanks}{\textsuperscript{\thefootnote}}

\begin{document}
\title{\textit{IRLCov19}: A Large COVID-19 Multilingual Twitter Dataset of Indian Regional Languages}

\author{Deepak Uniyal\inst{1} \thanks{Both authors contributed equally to this work.\protect\label{X}} \and
Amit Agarwal\inst{2} \repeatthanks}
\authorrunning{Deepak, Amit}

\institute{Graphic Era University, India\\
\email{deepak.uniya08@gmail.com}\\ 
\and
IIT Roorkee, India\\
\email{aagarwal3@cs.iitr.ac.in}
}
\maketitle              
\begin{abstract}
Emerged in Wuhan city of China in December 2019, COVID-19 continues to spread rapidly across the world despite authorities having made available a number of vaccines. While the coronavirus has been around for a significant period of time, people and authorities still feel the need for awareness due to the mutating nature of the virus and therefore varying symptoms and prevention strategies. People and authorities resort to social media platforms the most to share awareness information and voice out their opinions due to their massive outreach in spreading the word in practically no time. People use a number of languages to communicate over social media platforms based on their familiarity, language outreach, and availability on social media platforms. The entire world has been hit by the coronavirus and India is the second worst-hit country in terms of the number of active coronavirus cases. India, being a multilingual country, offers a great opportunity to study the outreach of various languages that have been actively used across social media platforms. In this study, we aim to study the dataset related to COVID-19 collected in the period between February 2020 to July 2020 specifically for regional languages in India. This could be helpful for the Government of India, various state governments, NGOs, researchers, and policymakers in studying different issues related to the pandemic. We found that English has been the mode of communication in over 64\% of tweets while as many as twelve regional languages in India account for approximately 4.77\% of tweets.

\end{abstract}

\section{Introduction}
\label{intro}
The novel coronavirus that erupted in December 2019 from Wuhan, China marked the beginning of the COVID-19 pandemic. With COVID-19 insurgence around the world, people are heavily dependent on social media platforms (SMPs) like Twitter to post their opinions, raise awareness among the general public, show their fear, ask for help, and communicate with fellow citizens.
Studies show that SMPs like Twitter has the potential to track emergencies in real-time that can be utilized by health officials, government agencies, and NGOs to respond quickly and more effectively\cite{broniatowski2013national}\cite{vieweg2010microblogging}.

Since the outbreak of the COVID-19 pandemic, most countries around the world have enforced several preventive and control measures to limit the spread of the virus. The measures range from early screening, isolation of patients, school and workplace closures, curfews, limited numbers of people in social gatherings, travel restrictions, social distancing to even complete lockdown in chosen cities or country as a whole\cite{guner2020covid}\cite{alqurashi2020large}. The success of these preventive measures would effectively imply people maintaining social distance as far as possible and use technology to interact and fulfill their day-to-day needs. In scenarios like this, SMPs like Twitter, Facebook, YouTube, Instagram, Snapchat, Reddit, Pinterest, and LinkedIn, etc. play a vital role by allowing individuals to interact thus helping them to alleviate social isolation. Contrary to the studies which suggest increased loneliness in people by excessive use of social media\cite{primack2017social}, SMPs have rather emerged as a friend to reduce isolation and boredom during the COVID-19 pandemic\cite{gonzalez2020social}.

Social media users may use a global or regional language to communicate on the platform based on their understanding of the language and ease of communication with other users producing a lot of data. With a plethora of unstructured data available on social media, it becomes crucial as to how one comprehends the information and uses it effectively to combat COVID-19.
India, the second-most populous country in the world, has 23 constitutionally recognized official languages which people may use to communicate. According to a census in 2001, Hindi is the most widely used language in India and is spoken by 53.6\% of the Indian population as their first language\cite{census2001}.

A major portion of the social media studies available today is based on the datasets in English. However, to better understand the information posted in the low-resource languages of the largest democracy in the world, we need to study the communication revolving around various Indian languages. Therefore, in this study, we have presented \textit{IRLCov19}, a large COVID-19 Twitter dataset on various Indian regional languages which we collected between 01 February 2020 to 31 July 2020\cite{Covid19IRLTDataset}. We collected nearly 330 million tweets irrespective of the language used and refined it further to remove tweets with duplicate IDs to make the final tweet count to 280 million. We subsequently identified more than 13 million tweets in twelve Indian Regional Languages (\textit{IRL}) from the dataset collected. This dataset can be advantageous for researchers, Government authorities, and policymakers in studying the pandemic from a varied perspective as listed below\cite{cha2010measuring}\cite{li2020characterizing}\cite{Prominent2021Disseminators}\cite{kouzy2020coronavirus} \cite{choi2020rumor}\cite{alharbi2021kawarith}:

\begin{itemize}
  \item \textbf{Public health strategies:} People post the situational information or content on social media corresponding to the need or availability of resources related to various emergency services such as medical supply, bed availability, blood or plasma donation, etc. The dataset we provided, can be used in developing suitable information publishing strategies by studying situational information to effectively respond in a pandemic situation.
  
  \item \textbf{Identification of echo chambers in social media:} Misinformation or rumors are said to be escalated by a group of users having similar ideologies or interests, known as an echo chamber of social media. This kind of dataset can be of great help in the identification and investigation of the characteristics or social properties of echo chambers which can be helpful in preventing rumor propagation in the early stage.
  
  \item \textbf{Understanding public reactions and opinions:} Public post their reactions, sentiments, and opinions on the various events, announcements, and actual implementation of fiscal and monetary policies initiated by the government during or after the pandemic. This kind of dataset can be used to study the pandemic from a social perspective, as well as analyzing the public opinions, human behavior, and information spreading pattern across the network.
  
  \item \textbf{Individual reaction on different policies roll-out by government:} The Reserve Bank of India, along with the government of India and other regulatory bodies, announced various fiscal and monetary measures to aid businesses during the lockdown. Several fiscal benefits by the government include cash transfers to lower-income households, wage support and employment provision to low-wage workers, and insurance coverage for workers in the healthcare sector etc. The monetary benefits include a reduction in the repo and the reverse repo rate by RBI. The government also announced several measures to ease the tax compliance burden such as postponing the tax and GST fillings. The analysis of the Twitter dataset can help gauze the public sentiment related to these policies. This would also help the government and authorities review how strategically the policies were implemented and were able to provide relief to the public.
  
  \item \textbf{Early detection and  surveillance of the pandemic:} Early detection of the pandemic can be helpful in preventing the further spread of the disease and loss of casualties. The analysis of Twitter data can help in the identification of content where masses may report their symptoms, reports, and localities, etc. which can be further used to identify the disease hot spots for prioritizing the further course of actions.
  
  \item \textbf{Identification of local or global leaders:} Identification of leaders or influencers is very significant during various emergency situations or natural disasters \cite{agarwal2020identifying} such as Covid-19, earthquake, glacier outbursts, floods, landslides, and wildfires, etc. because of their wide network, reach, popularity, or popular links. Such kind of users could use their remarkable network to spread the awareness information, debunk the misinformation or rumors as quickly as possible, ask for or provide help to the needy, communicate to authorities more effectively during the pandemic.

  \item \textbf{Tracking and debunking misinformation:} During critical and emergency situations it's of utmost importance to identify the misinformation, fake news, propaganda, or rumors and curb them as quickly as possible. It has been observed in the past studies that such kind of information spreads more quickly than the correct and factual information and therefore it becomes more important to identify and debunk such kind of unverifiable content that endangers public safety at a time when awareness and suitable preventive measures are of utmost importance and avoid any kind of panic in the public.
  
\end{itemize}

The rest of the paper is organized as follows. In the next section \ref{sec:related}, we describe COVID-19 related studies and datasets. In Section \ref{sec:dataCollection}, we provide the data collection and description in detail. Section \ref{sec:geoSpatial} is about geo-spatial analysis of tweets and section \ref{sec:influence} is about identification and analysis of user influence over the Twittersphere. Section \ref{sec:dataset} and section \ref{sec:conclusion} explains a way to access dataset and conclusion respectively.

\section{RELATED WORK}\label{sec:related}
There are a number of studies related to COVID-19 analysis of social media data being focused on various aspects such as human behavior and reactions analysis \cite{barkur2020sentiment}\cite{han2020using}, preparedness for emergency management\cite{li2020characterizing}, identifying and debunking conspiracy theories, misinformation, propaganda and fake news\cite{ferrara2020types}\cite{sharma2020covid}\cite{brennen2020types}\cite{gupta2020information}. Many other studies have collected and shared the COVID-19 related datasets from various social media platforms such as Twitter\cite{banda2020large}, Instagram\cite{zarei2020first}, Weibo\cite{hu2020weibo} etc. Some of the studies have released datasets belonging to single language such as Arabic\cite{alqurashi2020large}\cite{haouari2021arcov}, while others include multilingual datasets\cite{qazi2020geocov19}\cite{gao2020naist}\cite{aguilar2020dataset}\cite{shahi2020fakecovid}\cite{chen2020tracking}.

The largest available dataset contains 800  million tweets that are collected from 1 Jan 2020 to 8 Nov 2020 \cite{banda2020large}. The clean version of the dataset with no retweets is also provided which contains around 194  million tweets.
Another large dataset that is collected from 1st Feb 2020 to 1st May 2020 contains 524 million multilingual tweets\cite{qazi2020geocov19}. It also provides location information in the form of GPS coordinates and places information for some of the tweets as per the availability.
The longest-running dataset is of Arabic language\cite{haouari2021arcov} which is collected between 27 Jan 2020 to 31 Jan 2021. It also provides information related to propagation networks of the most-retweeted and most-liked tweets that include retweets and conversational threads i.e. threads of replies.
However, none of the above datasets focus on the \textit{IRL} and their research implications. We have included 12 Indian languages in our dataset, \textit{IRLCov19} which also includes location information with a subset of tweets depending on the availability of information. We have also analyzed the dataset to compute the local or regional influencers or leaders on the basis of various influencing measures such as \textit{followers}, \textit{retweet count}, \textit{favourite count} and number of \textit{mentions}, which is discussed in detail in section \ref{sec:influence}.

\begin{table*}[t]
\caption{Language Wise Tweets Distribution}
	\label{table:languageDistribution}
\centering
\resizebox{\linewidth}!{
\begin{tabular}{|l|l|l|l|l|l|l|l|} \hline
\textbf{Language} & \textbf{Percentage} & \textbf{Language} & \textbf{Percentage} & \textbf{Language} & \textbf{Percentage} & \textbf{Language} & \textbf{Percentage} \\ \hline
English & 64.11 & \textbf{Marathi} & 0.19 & Danish  & 0.018 & Latvian  & 0.003 \\ \hline
Spanish & 14.08 & Greek & 0.14  & \textbf{Malayalam}  & 0.017 & \textbf{Sindhi}  & 0.003 \\ \hline
French   & 5.003  & \textbf{Telugu}  & 0.11 & Swedish & 0.017 & Hebrew  & 0.002 \\ \hline
\textbf{Hindi} & 3.36  & Chinese  & 0.101  & Finnish & 0.017 & Maldivian  & 0.001 \\ \hline
Italian  & 2.1  & Tagalog  & 0.09  & Basque  & 0.0127  & Amharic  & 0.001  \\ \hline
Thai & 1.8  & Polish & 0.09  & Slovenian  & 0.012 & Icelandic  & 0.001 \\ \hline
Undefined & 1.76  & \textbf{Gujarati} & 0.071  & Czech & 0.0106 & Bulgarian  & 0.001 \\ \hline
Portuguese & 1.45  & Persian  & 0.07  & \textbf{Punjabi}  & 0.01  & Sorani Kurdish  & 0.001 \\ \hline
German   & 0.96 & \textbf{Kannada}  & 0.059  & Sinhala  & 0.01 & Armenian & 0.0001 \\ \hline
Turkish  & 0.87 & Russian  & 0.05 & Ukrainian  & 0.007  & Burmese  & 0.0001  \\ \hline
Indonesian  & 0.71 & Estonian & 0.04 & Welsh & 0.006   & Georgian & 0.00005 \\ \hline
\textbf{Tamil} & 0.55 & \textbf{Bengali} & 0.028 & Serbian & 0.005 & Khmer  & 0.00004  \\ \hline
Catalan  & 0.5 & Haitian Creole  & 0.025  & Lithuanian & 0.005 & Laotian Lao  & 0.00004 \\ \hline
Arabic  & 0.42 & Romanian & 0.023  & Norwegian  & 0.005  & Uyghur  & 0.00003  \\ \hline
\textbf{Urdu} & 0.36 & Korean   & 0.02  & Hungarian  & 0.005 & Tibetan  & 0.00002 \\ \hline
Dutch & 0.33 & \textbf{Oriya}  & 0.02  & Pashto  & 0.005  & &  \\ \hline
Japanese & 0.29  & Nepali   & 0.019 & Vietnamese & 0.003  &   & \\ \hline
\end{tabular}}
\end{table*}

\section{DATA COLLECTION AND DESCRIPTION}\label{sec:dataCollection}
We collected Twitter datasets on COVID-19 during the period from Feb 01 2020 to July 31 2020 using publicly available Twitter streaming API. To download the dataset we utilized a list of trending keywords and hashtags such as \textit{corona}, \textit{Covid-19}, \textit{\#COVID19}, \textit{\#COVID2019}, \textit{\#Covid\_19}, \textit{\#CoronaVirusUpdates} etc. We kept updating the list of keywords and hashtags as and when they were available daily. 

Initially, we collected a  dataset of nearly 330 million tweets irrespective of the language of communication. The \textbf{Table} \ref{table:languageDistribution} gives the percentage-wise distribution of tweets collected between a given time period. The downloaded tweets may be redundant as a tweet may contain multiple search keywords and therefore get downloaded multiple times for each such keyword. It is imperative to remove such occurrences for a more robust dataset.  We pruned the dataset to remove the redundant tweets to result in over 280 million final tweets. We extracted the tweets specific to 12 Indian languages marked in bold in \textbf{Table} \ref{table:languageDistribution}. Owing to a small percentage, we could infer from the dataset that not many people were using regional languages for communicating on Twitter. Another reason for this could be that the hashtags or mentioned used by regional languages' users could not find a place in the trending list of keywords. We have utilized trending hashtags or keywords and hence the latter could be a strong possibility. It is evident from the dataset that people have used various global, national or regional languages to voice out their opinions on varying matters. English comprises 64.11\% of the total tweets out of all 65 languages in the dataset.

\begin{figure*}[t]
\centering
\includegraphics[width=\linewidth]{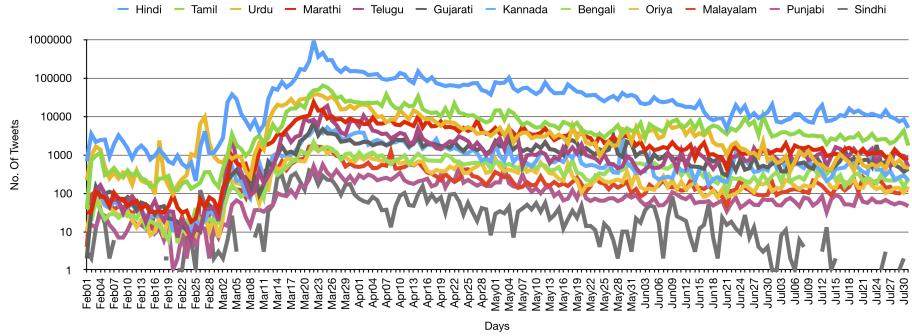}
\caption{Day Wise Tweets Distribution for Indian Regional Languages on Logarithmic Scale}
\label{fig:dailytweetCount}
\end{figure*}

In this study, we have focussed on studying the dataset on \textit{IRL} that constitute approximately 13 Million (1,33,63,294) tweets which are about 4.77\% of the total collected tweets. The daily distribution of the tweets corresponding to various regional languages is shown in \textbf{Fig. }\ref{fig:dailytweetCount} on a logarithmic scale. It represents the volume of tweets against each language for a period of six months starting Feb 01 2020 to July 31 2020. The data in the table shows that tweets in the Hindi language are consistently high in numbers compared to other \textit{IRL}. The findings coincide with the fact that Hindi is the most spoken language in the country. The high spikes in the graph after mid-March mark the beginning of a voluntary public curfew on March 22 2020. As evident from the plot, this was followed by a 21-day nationwide lockdown starting from March 25 2020, which resulted in masses expressing themselves on various SMPs. \textbf{Table }\ref{table:allUsers} has the count of users with original as well as re-tweeted tweets. The data shows that a total of 14,28,876 unique users were involved in exchanging thoughts and opinions in \textit{IRL}. While most of these users are non-verified users, a little less than 1\% are verified.

The dataset prepared is for research and non-profit uses and includes keywords used for dataset extraction, unique tweet IDs, and everyday language-wise tweet count. We first removed all duplicate entries by keeping the first instance of a tweet and kept a list of tweets corresponding to each language for all days. Later, we identified the location information from each tweet using the metadata in a tweet.

\textbf{Identification of Location Information From Tweets}
The location information in a tweet can be identified in three different ways i.e by extracting the location information in the form of GPS coordinates from the downloaded JSON Twitter data, using the place or location information from the Tweet object\cite{uniyal2019citizens}\cite{uniyal2020social}\cite{agarwal2018geospatial} and extracting the location information from the textual content\cite{agarwal2019face}. The previous studies show that only 1\% of the tweets contain GPS coordinates despite Twitter providing an option to capture the exact location of Tweet by enabling geolocation service on mobile devices. We can also deduce the location from the place and location fields in the JSON data. It can correctly identify the approximate location of the user but not the location of the tweet in all cases. Not all the values in these fields are valid locations; for example, \textit{Universe, Moon, Planet Earth, Heaven etc}. Invalid locations can be handled by transforming them to coordinates i.e. latitude and longitude, by using a python library called GeoPy. It returns coordinates for only valid locations by discarding invalid locations. The library may not always correctly classify valid locations due to misspellings or other possible errors in the text. This inhibits its ability to correctly map the geocode from a given location and such scenarios have been handled manually. The location of a tweet can also be deduced by exploiting the information in the text or by looking at its network of followers or friends. This method of capturing locations can be explored further in future works. 

In this study, we have used two parsers i.e. $P1$ and $P2$ to extract the location information from Tweets. The extracted location could be in terms of the GPS coordinates or place and location fields as extracted from the profile information. Parser ($P1$) looks for the geo-coordinates that comprise both latitude and longitude in a tweet. Parser ($P2$), however, extracts the place and location fields from the JSON data of the tweet in case ($P1$) could not find geo-coordinates.
A retweet contains the profile information of both the original user, known as the source of the tweet,  and of the user who retweeted it. Parser $P2$ prioritizes the information of retweeting users over original users while searching for a place or location information as multiple retweets are possible for a tweet across the globe.

\begin{table}[t]
\caption{Users Details Corresponding To Indian Regional Languages (\textit{IRLCov19})}
	\label{table:allUsers}
\centering
\resizebox{0.7\linewidth}!{
\begin{tabular}{|l|l|l|l|l|} \hline

\diagbox{\textbf{Unique}}{\textbf{Count}}   & \textbf{Verified} & \textbf{Non-Verified} & \textbf{Total} & \textbf{Verified \%} \\ \hline
Original Users & 3498     & 437339       & 440837         & 0.79\%                         \\ \hline
Retweet Users  & 4146	     & 1178347      & 1182493       & 0.35\%                         \\ \hline
All Users      & 5284     & 1423592      & 1428876        & 0.37\%                        
\\\hline
\end{tabular}}
\end{table}

\begin{figure}[t]
	\centering
	\subfloat[Hindi]
	{\includegraphics[width=2.5cm,height=2.5cm]{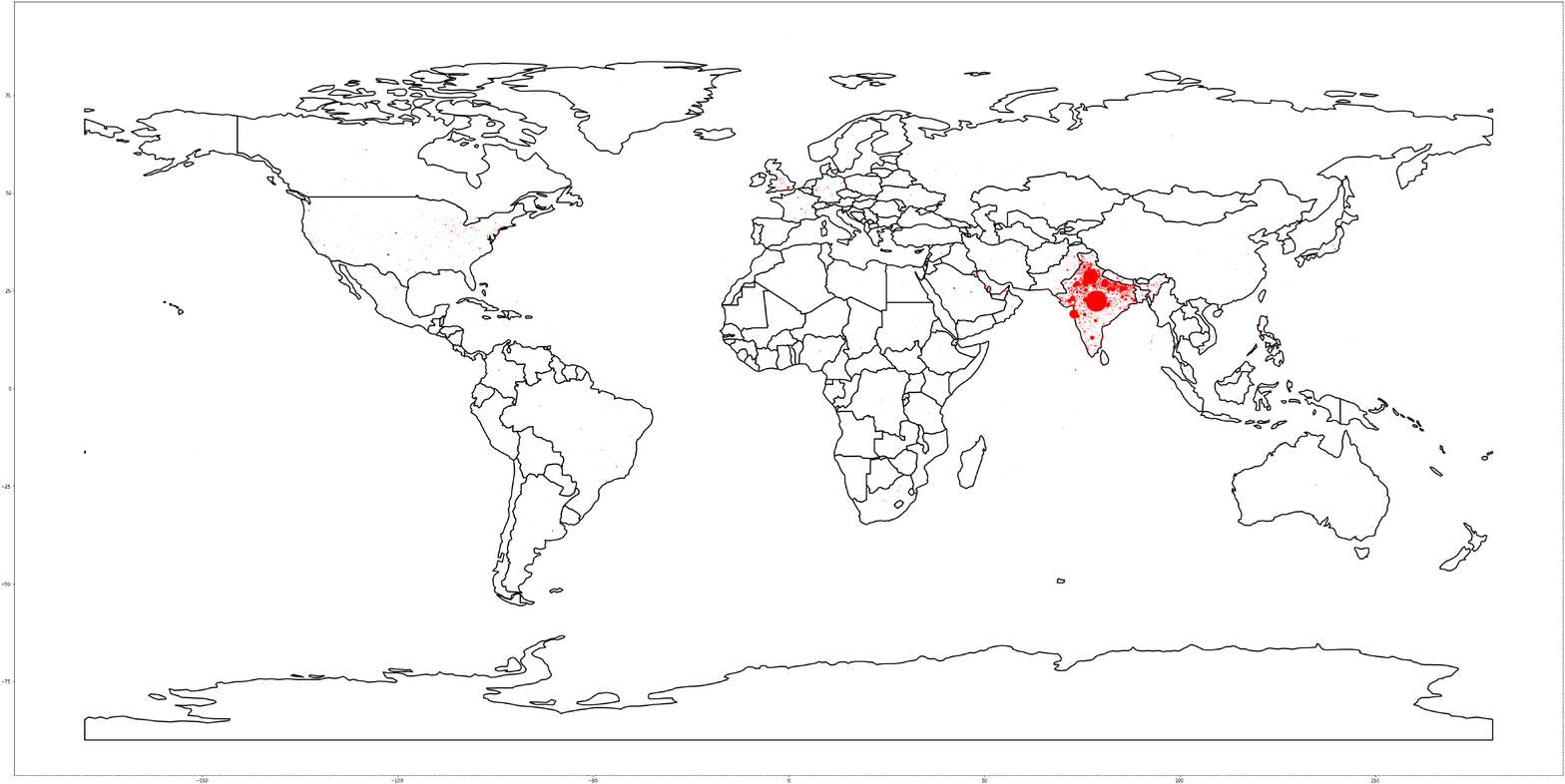}%
		\hspace{1mm}
		\label{f1}}
	\subfloat[Tamil]
	{\includegraphics[width=2.5cm,height=2.5cm]{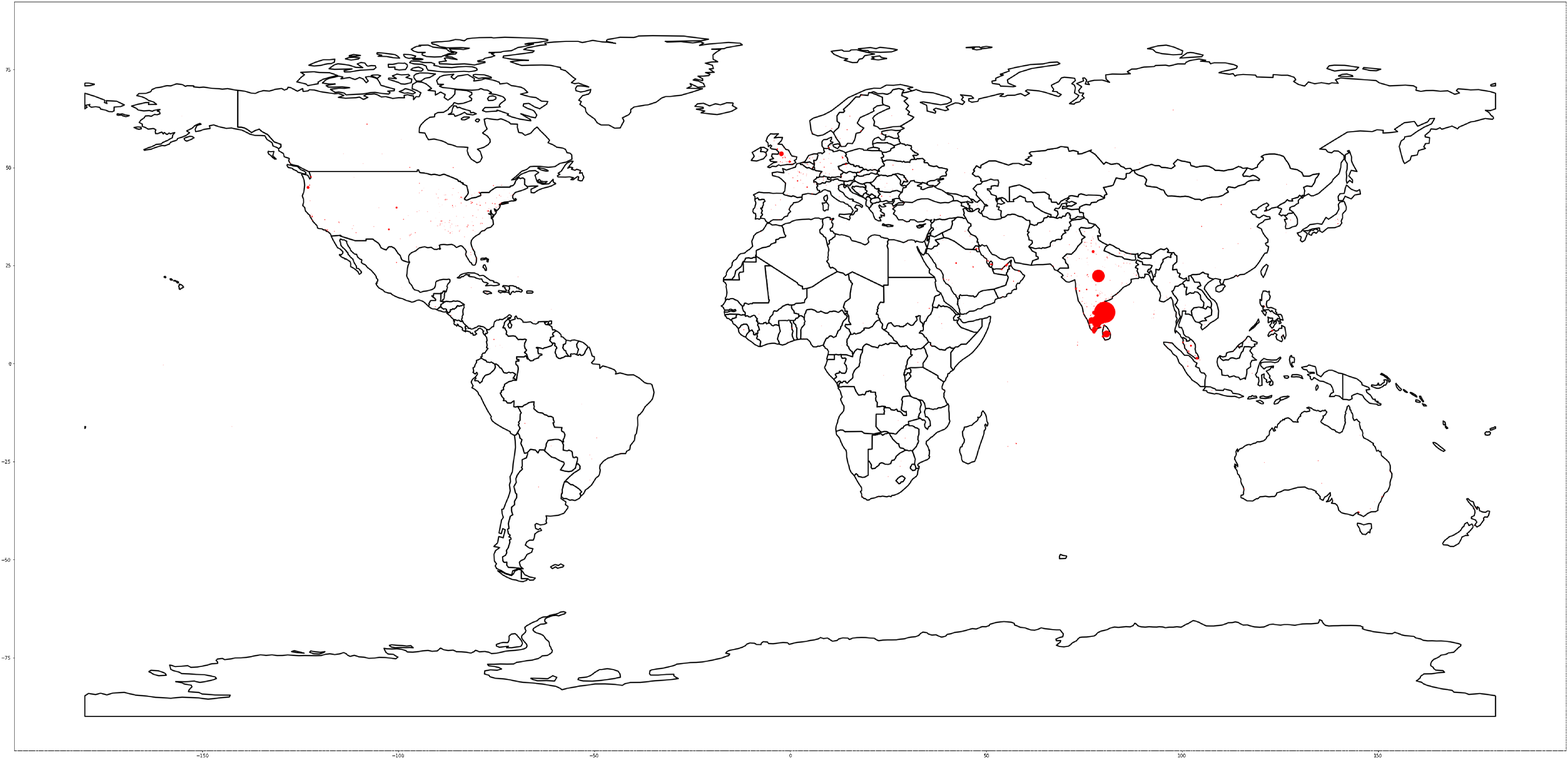}%
			\hspace{1mm}
		\label{f2}}
	\subfloat[Urdu]
	{\includegraphics[width=2.5cm,height=2.5cm]{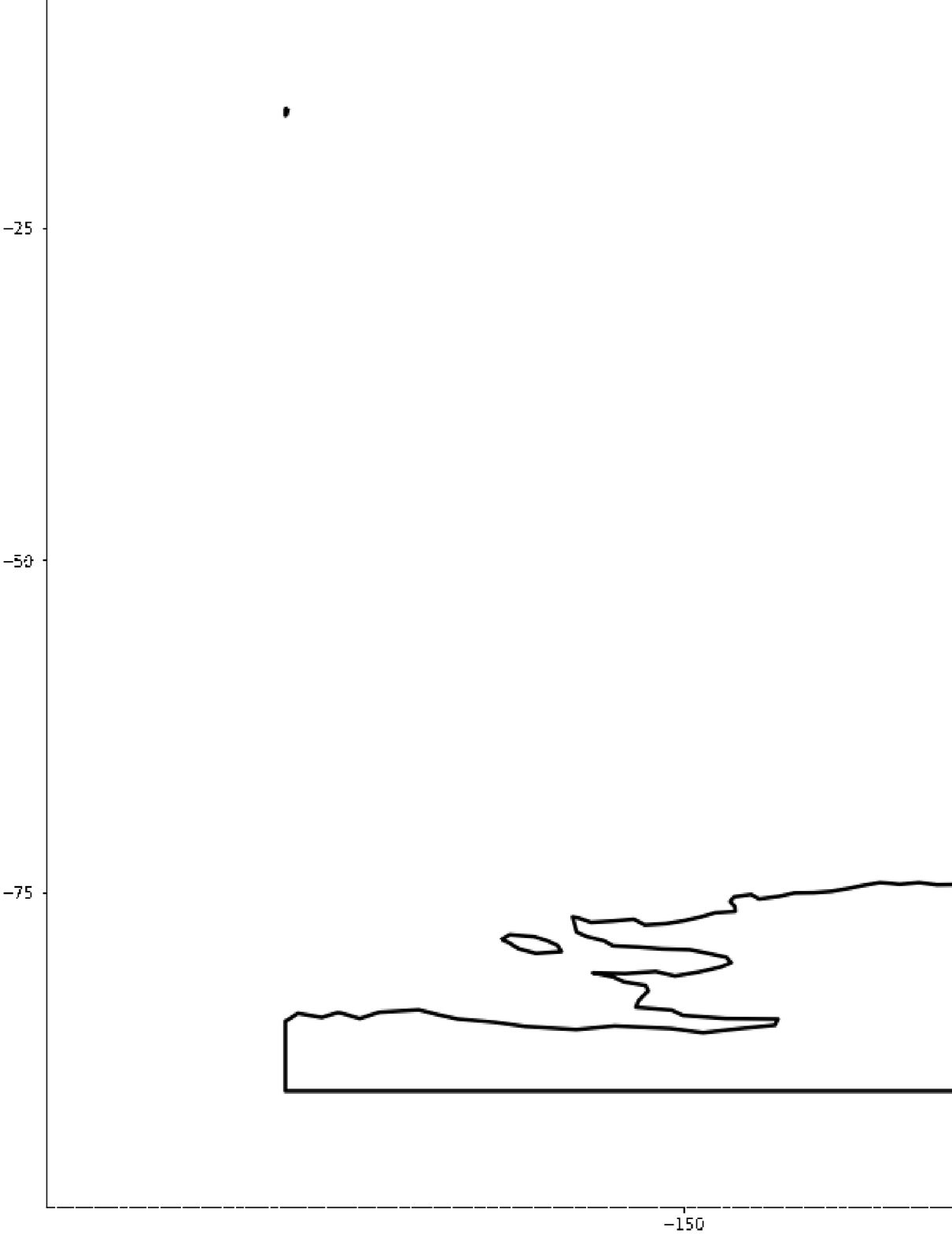}%
	  \hspace{1mm}
		\label{f3}}
	\hspace{01mm}
	\subfloat[Marathi]
	{\includegraphics[width=2.5cm,height=2.5cm]{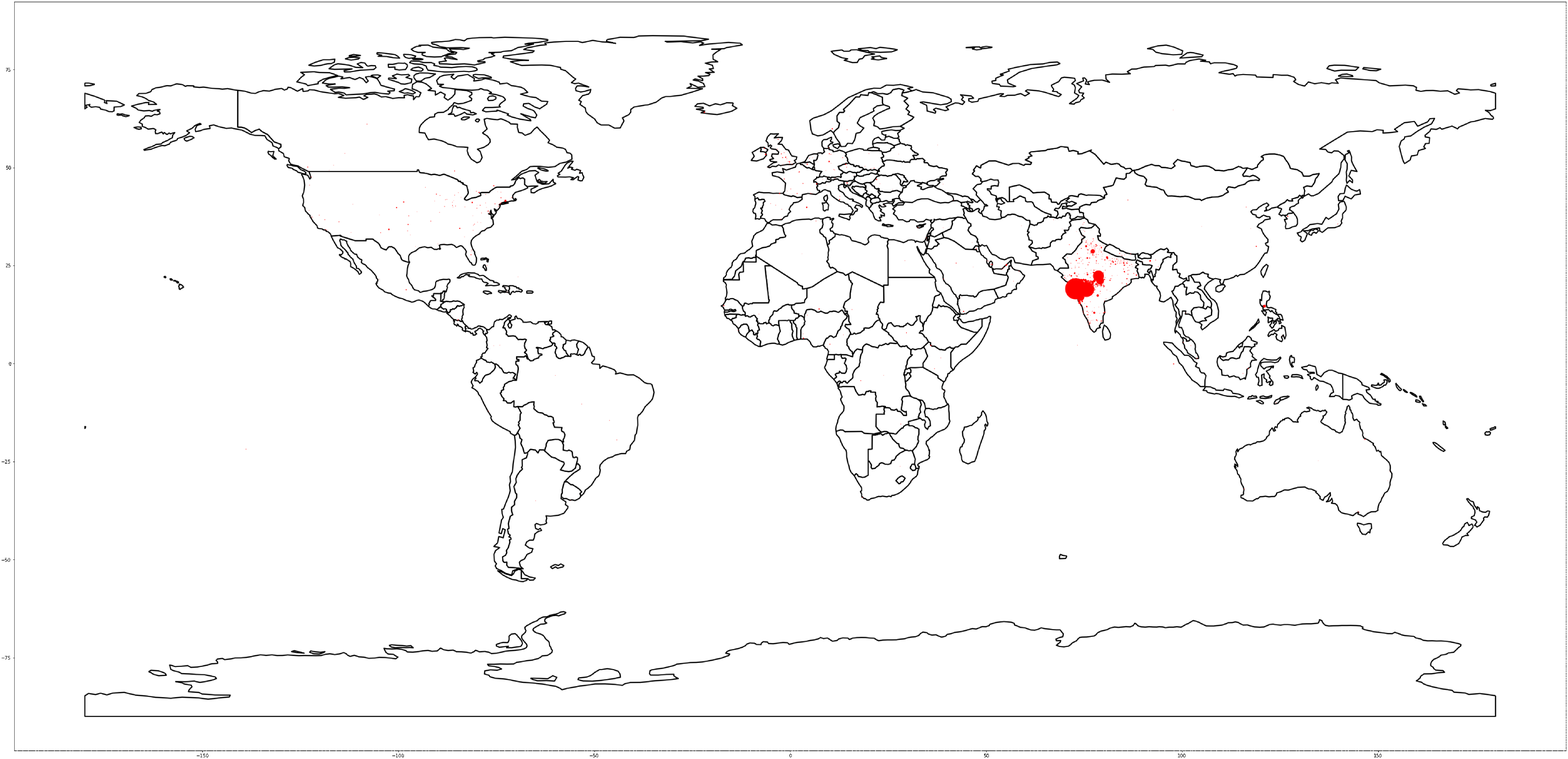}%
	  \hspace{1mm}
		\label{f4}}\\
	\subfloat[Telugu]
	{\includegraphics[width=2.5cm,height=2.5cm]{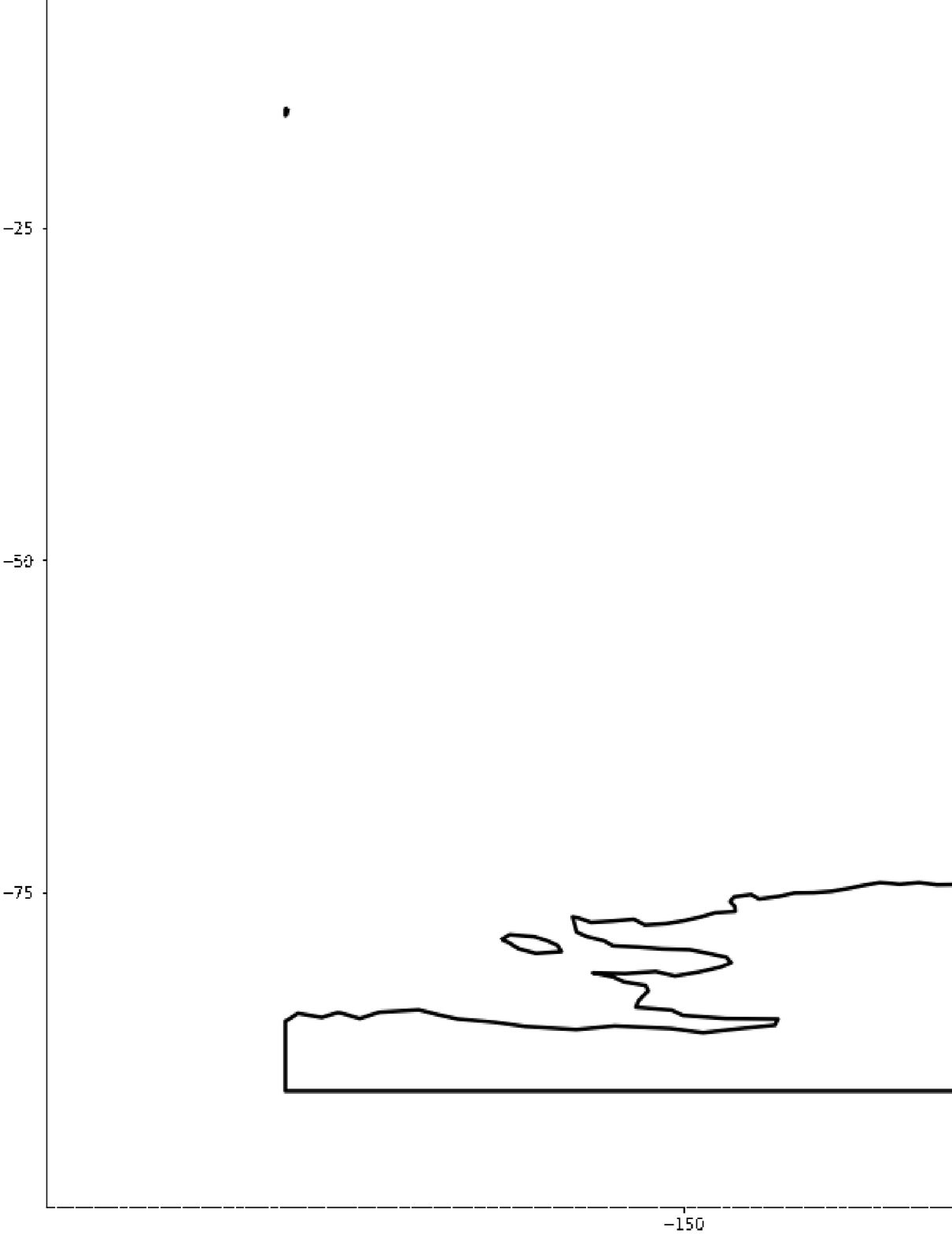}%
		\label{f5}}
		\hspace{1mm}
			\subfloat[Gujarati]
	{\includegraphics[width=2.5cm,height=2.5cm]{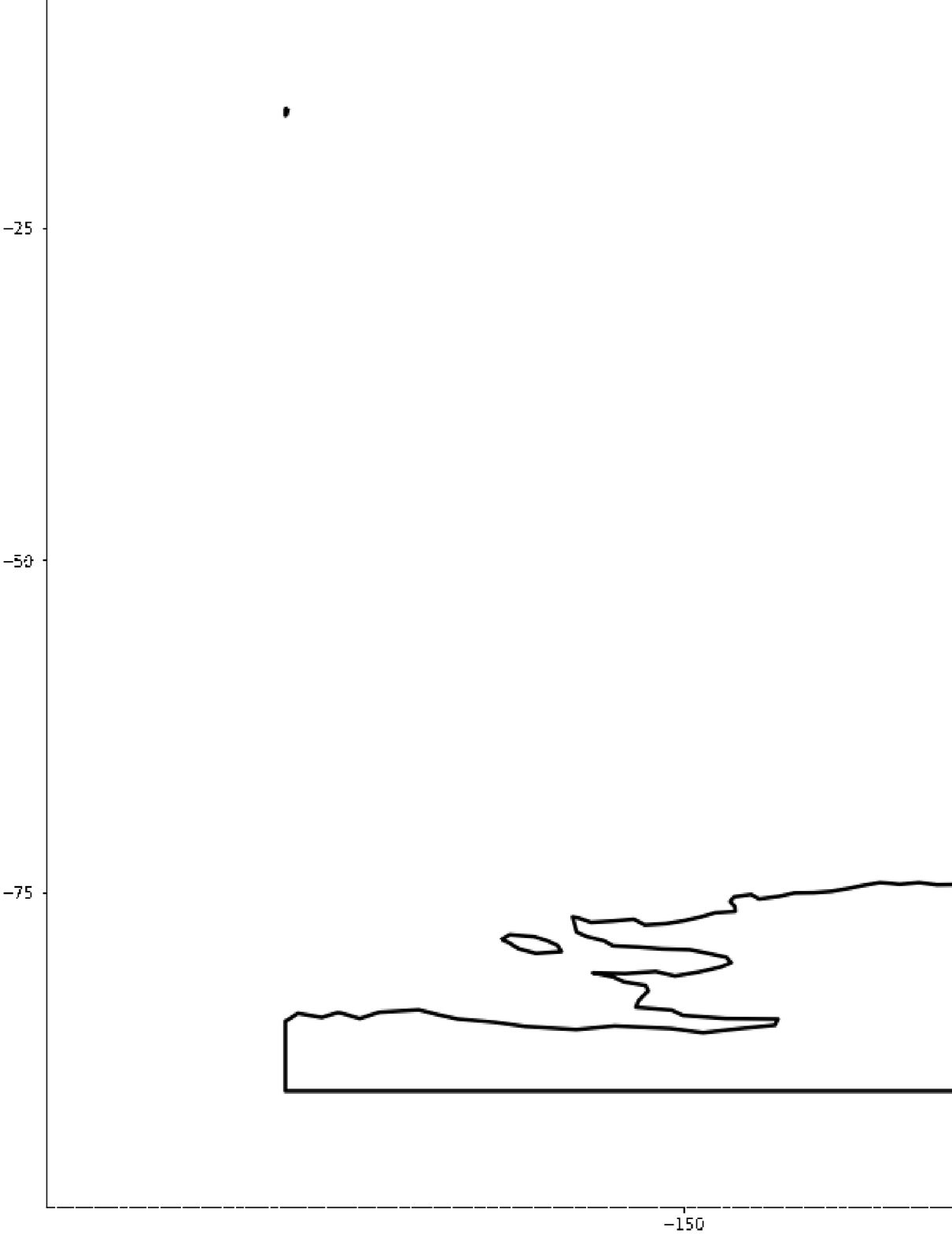}%
		\label{f6}}
		\hspace{1mm}
		\subfloat[Kannada]
	{\includegraphics[width=2.5cm,height=2.5cm]{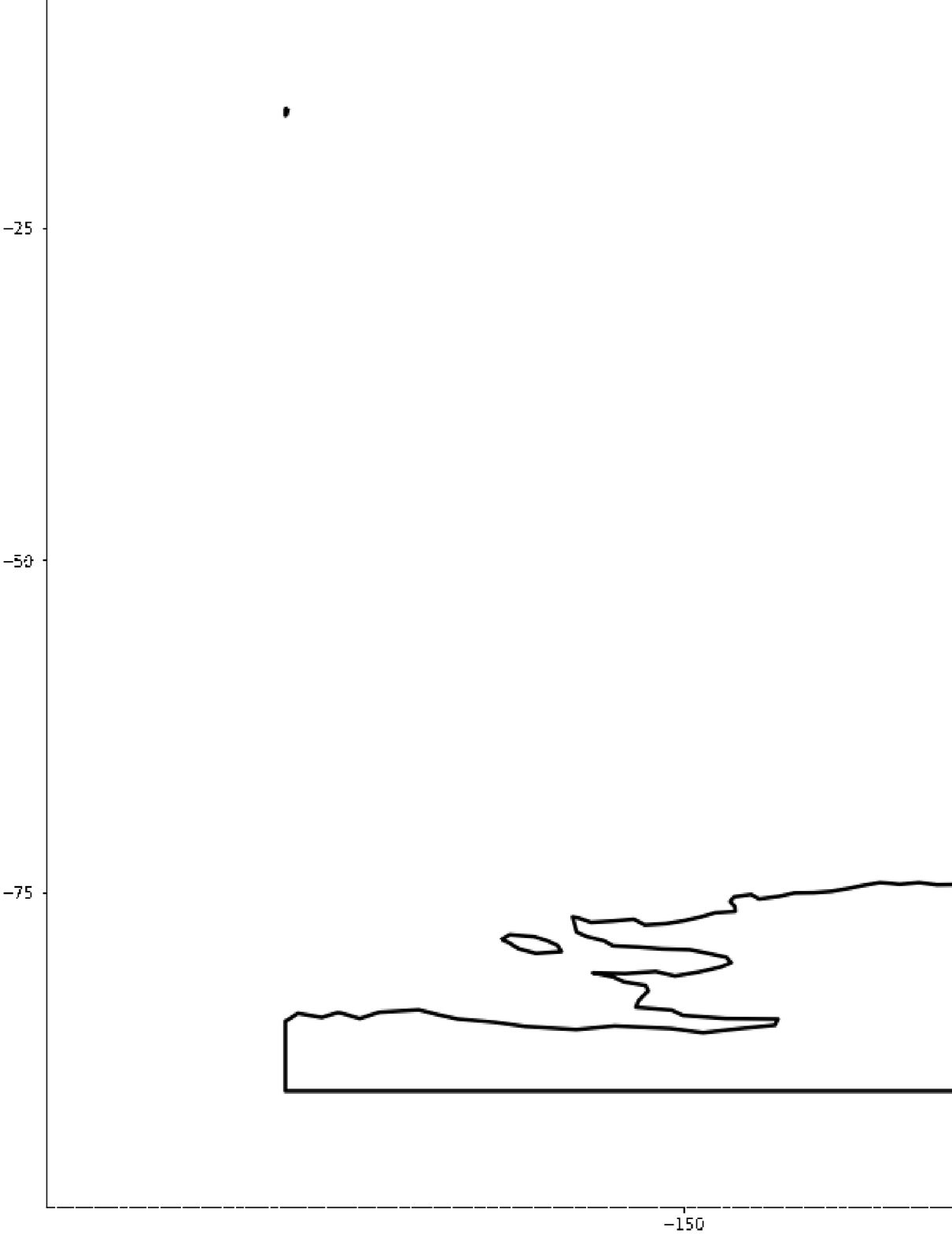}%
	  \hspace{1mm}
		\label{f7}}
	\subfloat[Bengali]
	{\includegraphics[width=2.5cm,height=2.5cm]{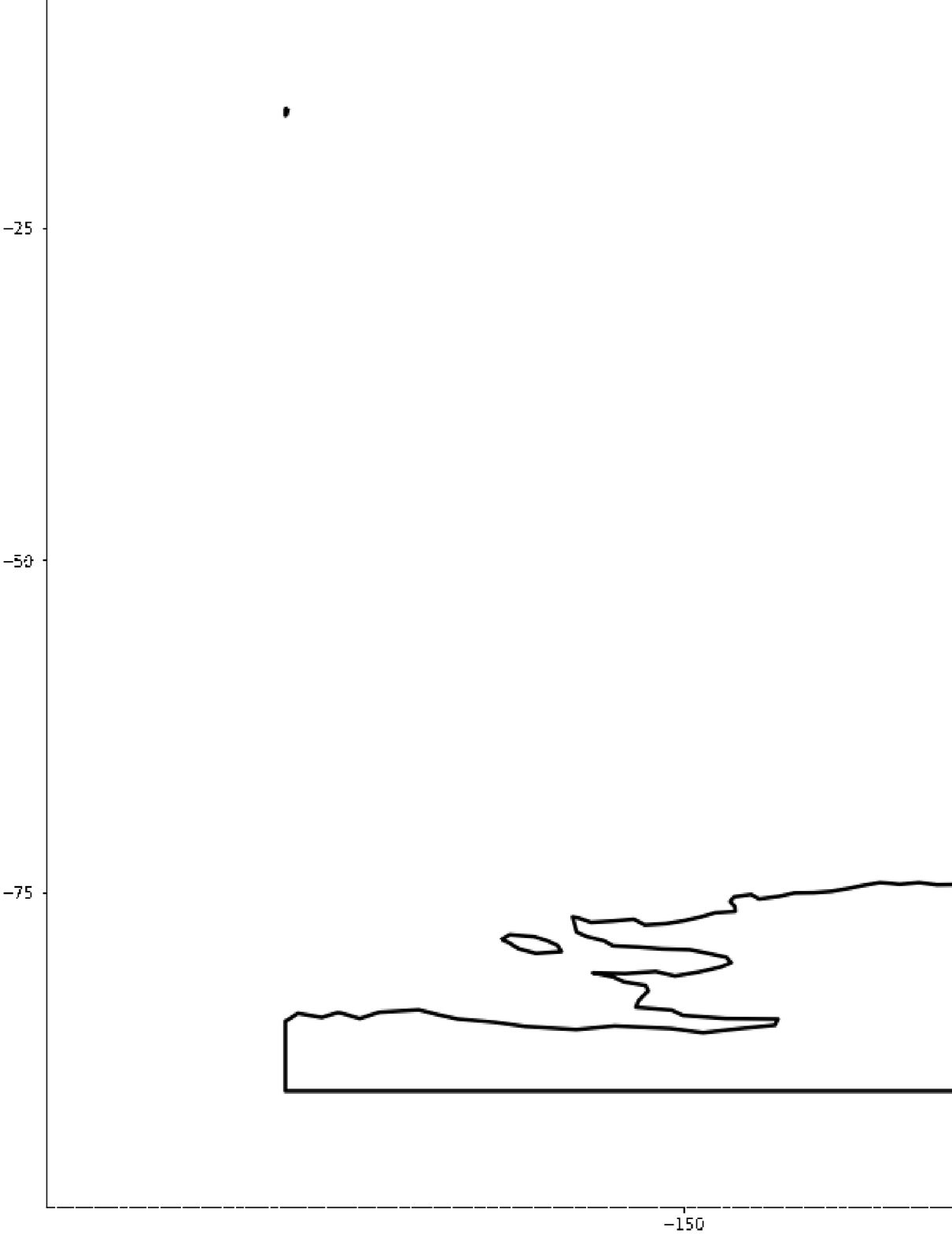}%
		\hspace{1mm}
		\label{f8}}\\
	\subfloat[Malayalam]
	{\includegraphics[width=2.5cm,height=2.5cm]{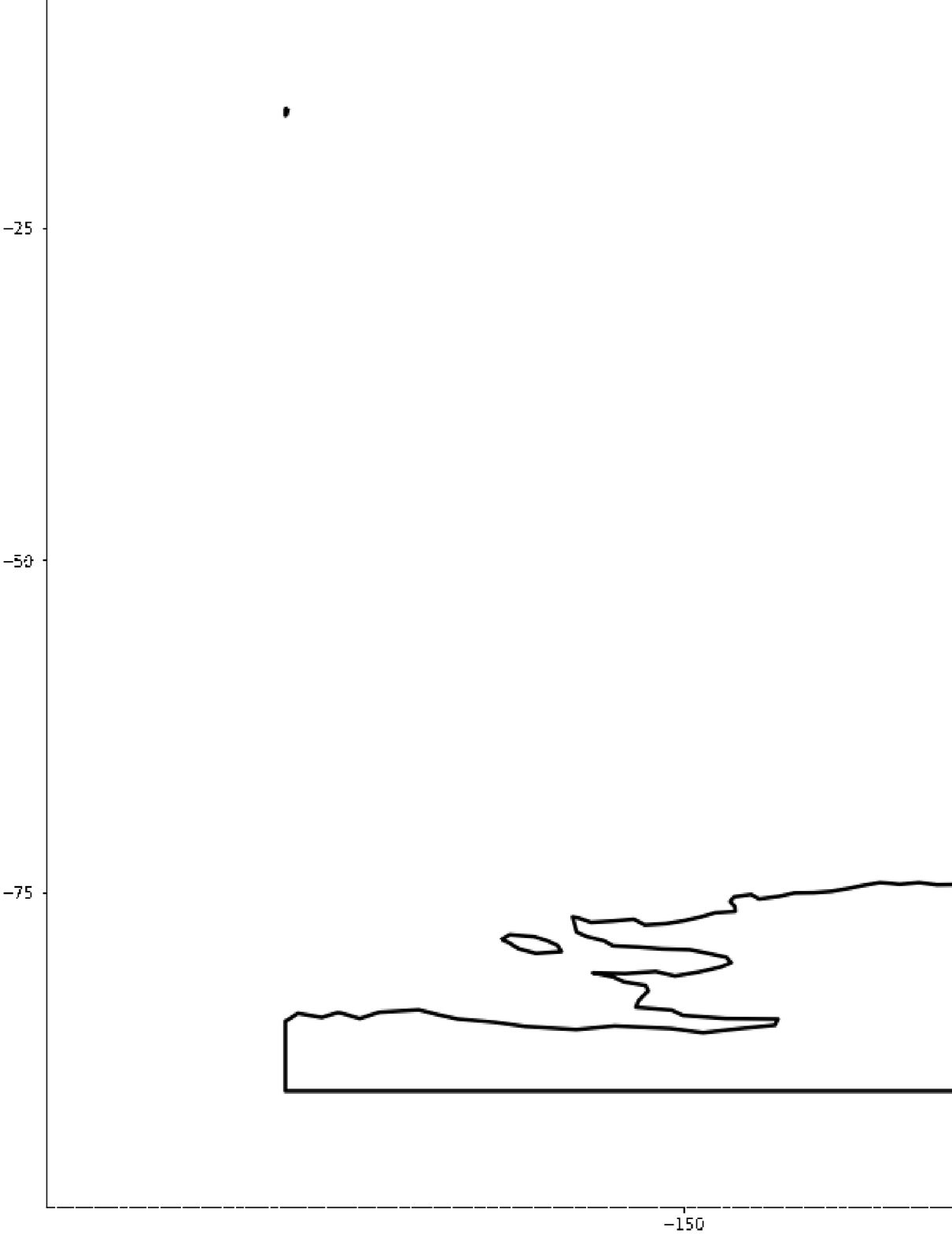}%
		\hspace{1mm}
		\label{f9}}
	\subfloat[Oriya]
	{\includegraphics[width=2.5cm,height=2.5cm]{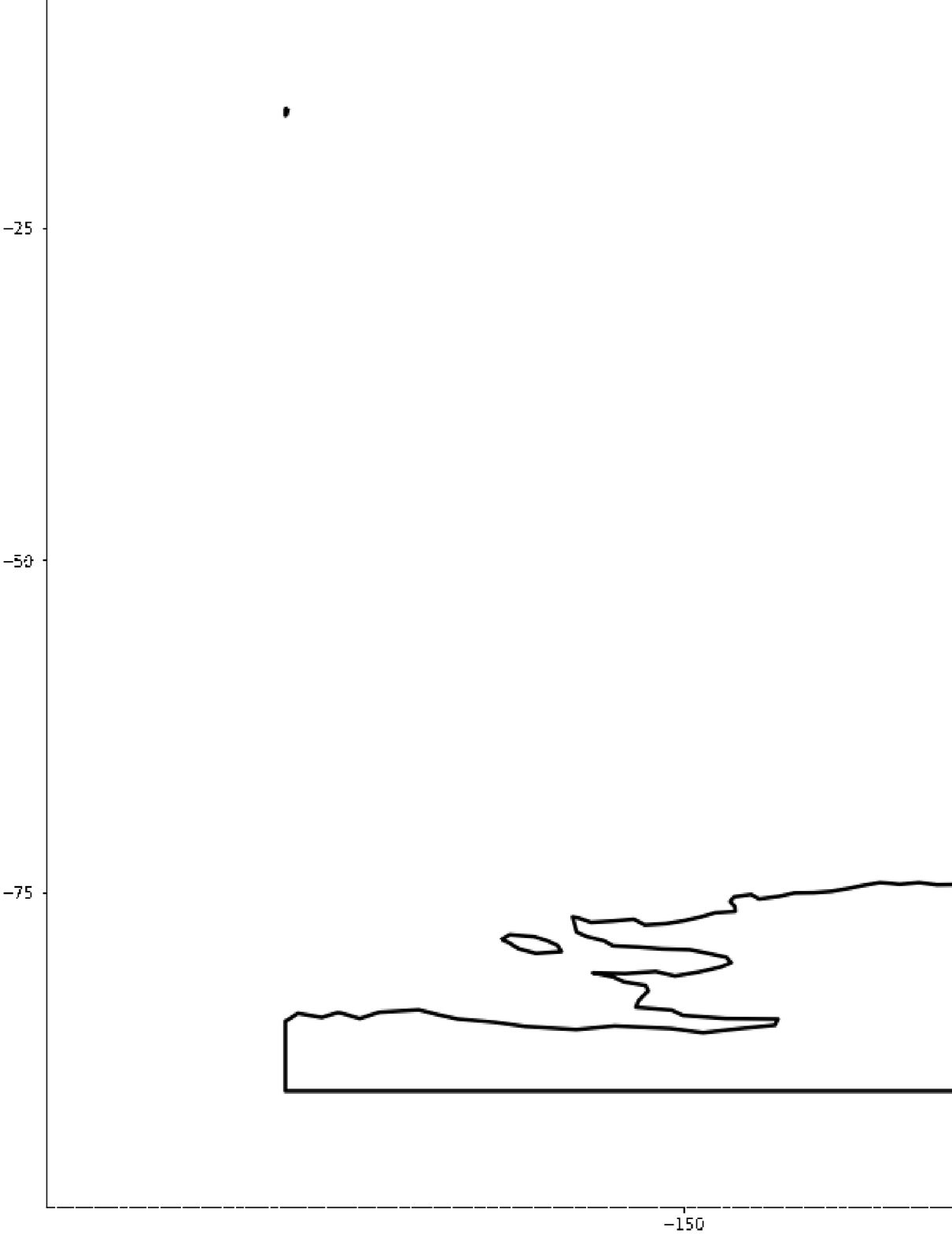}%
	  \hspace{1mm}
		\label{f10}}
		\subfloat[Punjabi]
	{\includegraphics[width=2.5cm,height=2.5cm]{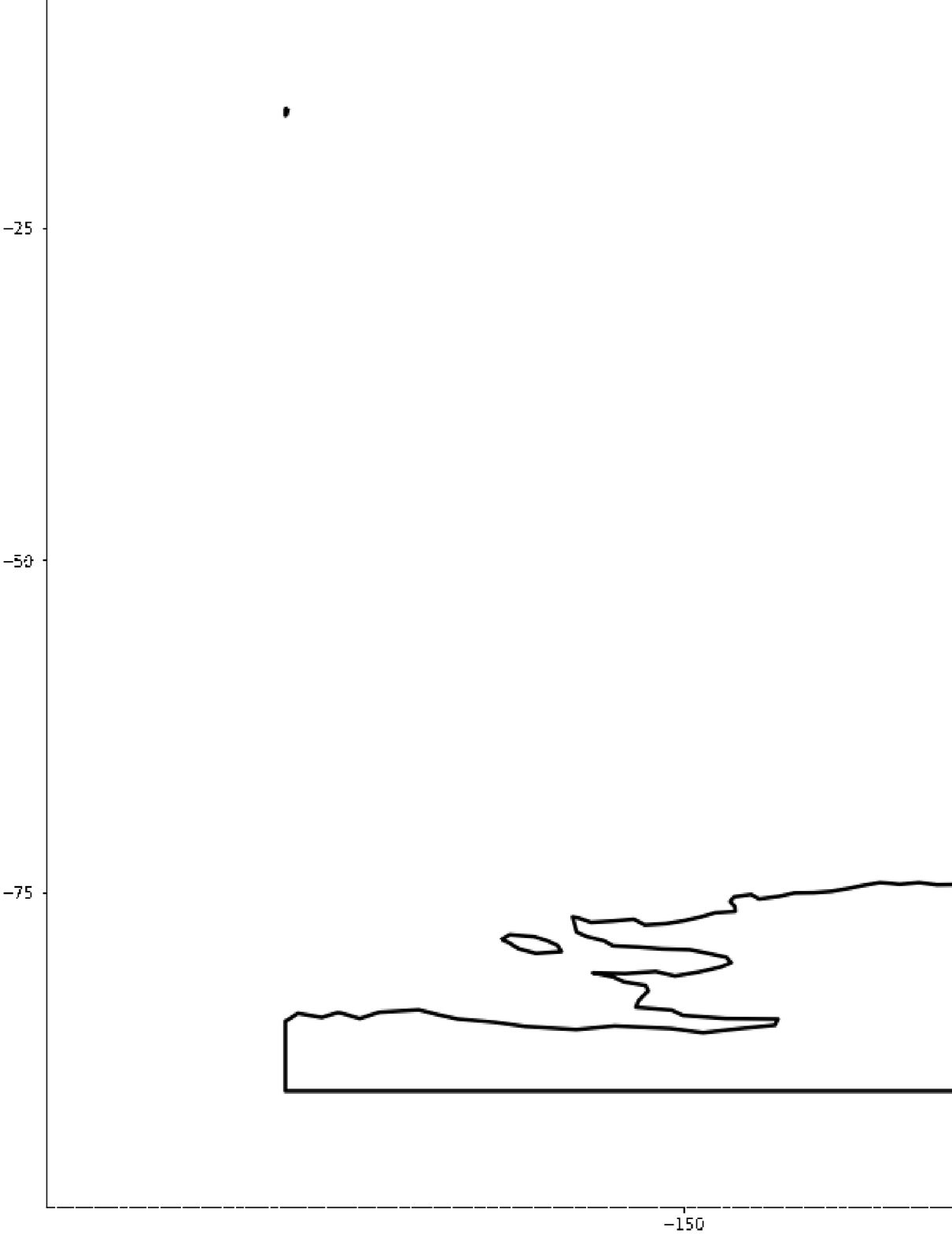}%
		\hspace{1mm}
		\label{f11}}
	\subfloat[Sindhi]
	{\includegraphics[width=2.5cm,height=2.5cm]{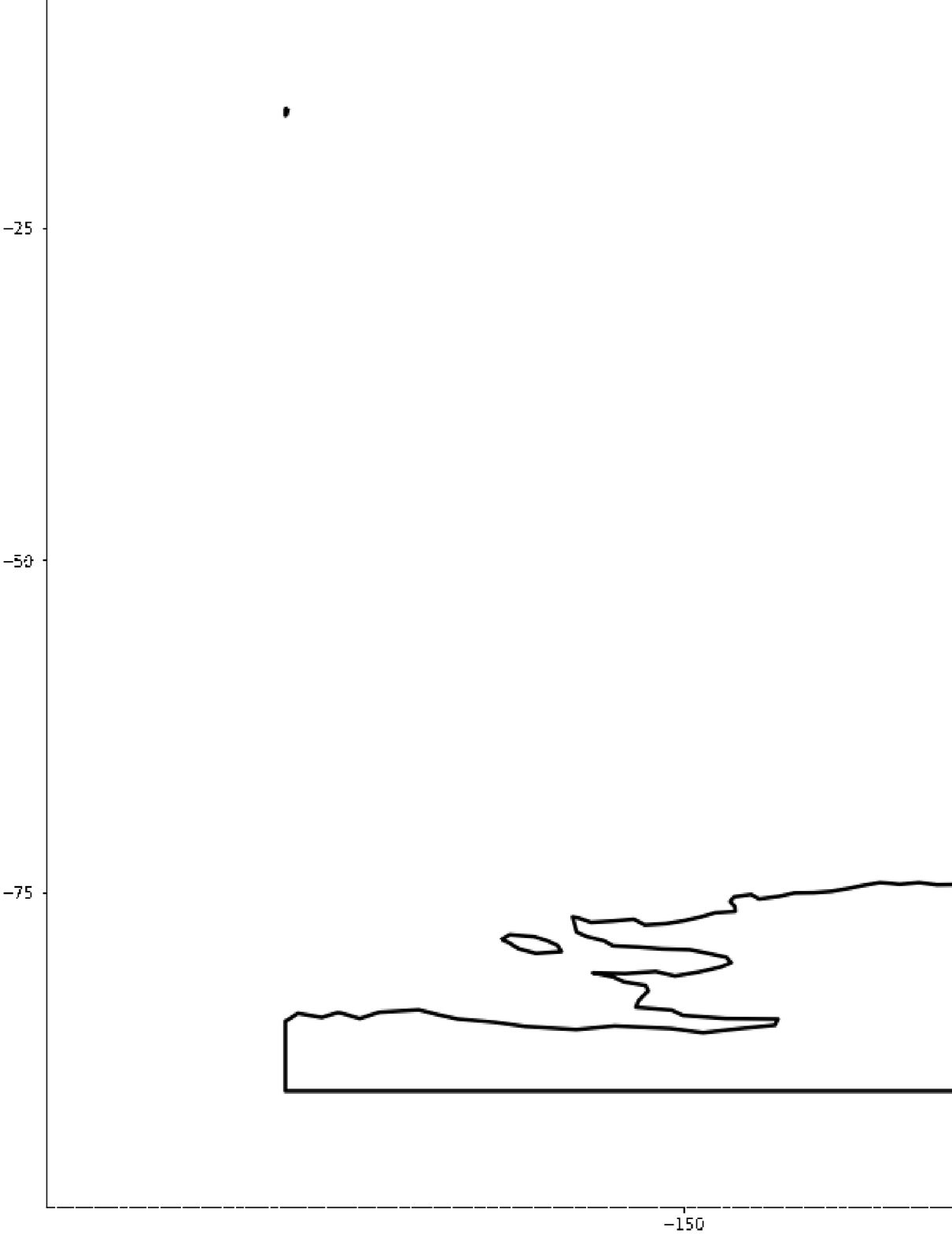}%
		\hspace{1mm}
		\label{f12}}\\

\caption{Visualization of Locations from Indian Regional Languages Tweets}
\label{fig:locations}
\end{figure}

\begin{figure}[t]
\centering
\includegraphics[width=0.7\linewidth]{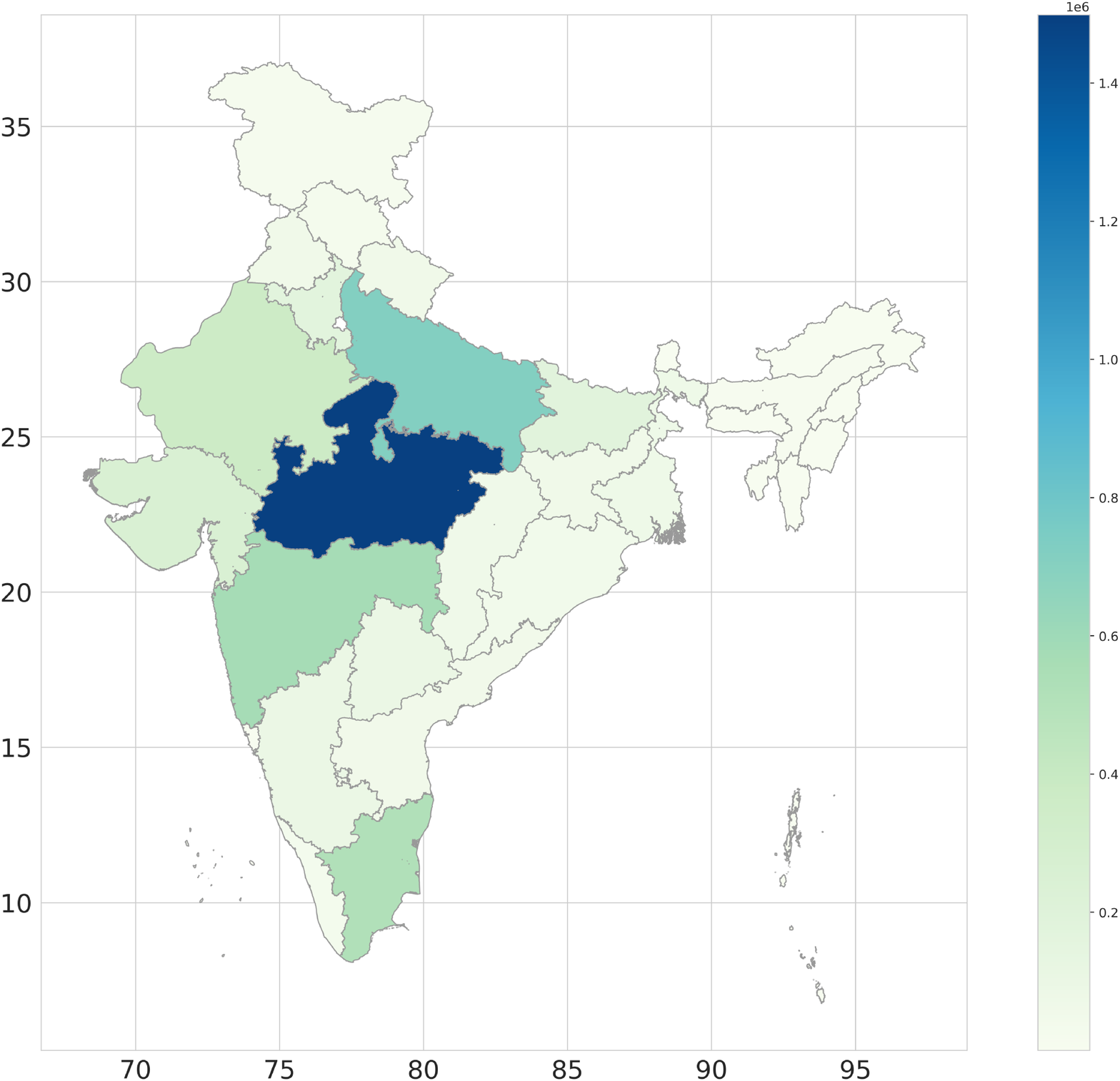}
\caption{State Wise Tweets Distribution}
\label{fig:stateWiseTweetsDistribution}
\end{figure}

\begin{figure}[t]
\centering
\includegraphics[width=0.8\linewidth]{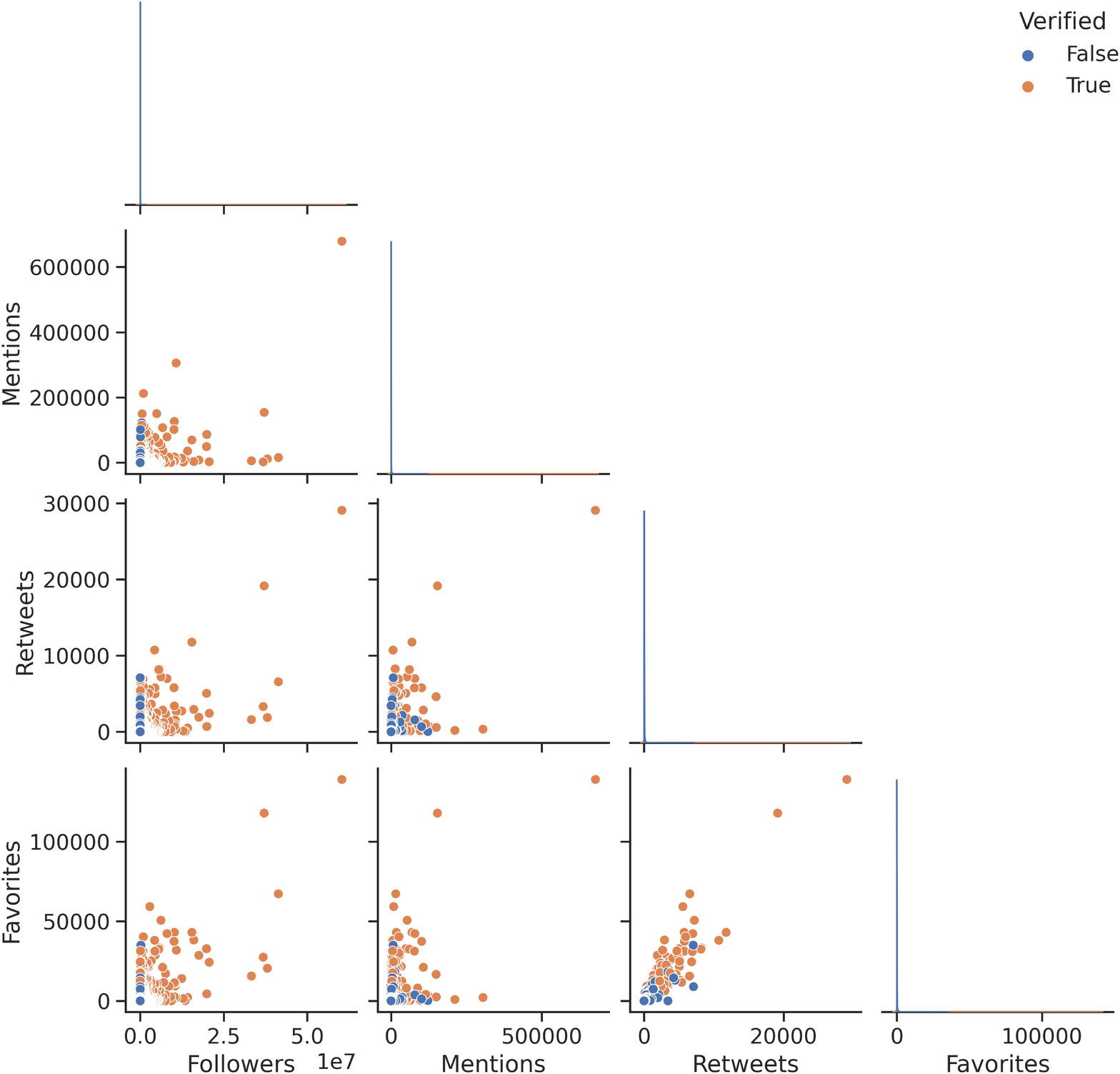}
\caption{Scatterplots of Influencing Variables}
\label{fig:scatterplots}
\end{figure}

\section{Geo-Spatial Analysis of Tweets}\label{sec:geoSpatial}
Geo-Spatial analysis of tweets during emergencies, such as pandemics and natural disasters, plays a vital role in identifying the pattern of information propagation in the affected areas of the leaders involved in the communication. Therefore, the information gathered can be helpful for various regional, national, and global organizations to evaluate the circumstances and develop a strategy to combat the crises. Notably, it can be utilized to identify the prominent leaders around a region working as spreaders of information or misinformation on the social network. This could essentially be used in numerous ways like disseminating the awareness information, communicating the policies or schemes launched by Governments, reaching out to needy people promptly, and tracking down the source of misinformation to put measures in place, etc.

To analyse the locations of \textit{IRLCov19} dataset with respect to each language, we have transformed identified valid locations to their corresponding geo-coordinates with the help of GeoPy\cite{geopy} - A Python client for several popular geocoding web services, that includes geocoder classes for the \textit{OpenStreetMap Nominatim}, \textit{Google Geocoding API (V3)}, and many other geocoding services. We have used \textit{Nominatim}, for:\\
\textbf{1.} \textit{Geocoding} - the process of obtaining GPS coordinates cooresponding to a valid location. \\ \textbf{2. }\textit{Reverse Geocoding} - the process of obtaining location names using GPS coordinates.

Firstly, we listed all the locations and coordinates along with their number of occurrences corresponding to each language.Given that some of these locations were already in the form of coordinates, we transformed the remaining locations available in the textual format into coordinates using geocoding service of \textit{Nominatim}. We plotted the coordinates over a world map where the size of a \textcolor{red}{RED} dot is directly proportional to the frequency or number of tweets done around that location as shown in \textbf{Fig. }\ref{fig:locations}. The map shows that most of the tweets corresponding to each language have originated from the region of the country where the language is accepted as a regional language. In certain cases, tweets of an otherwise regional language could also originate from a location around the globe based on where the users of the language reside.

Tweets that mention India as their location are by default mapped to the common latitude and longitude coordinates \textit{22.3511148} , \textit{78.6677428} and can be seen marked in red in most of the maps. The distribution of state-wise tweets is also shown in \textbf{Fig. }\ref{fig:stateWiseTweetsDistribution}. The colour intensity shows the frequency of tweets with the maximum intensity denoting the highest frequency.

\section{Identification and Analysis of User Influence Over The Twittersphere}\label{sec:influence} 
There are multiple ways to identify a user's influence over the network, such as In-degree (number of followers), number of retweets, number of favourites (likes) received by the user on tweets, and number of mentions for the user in related discussions over a period of time. The metrics, \textit{In-degree}, \textit{retweets}, \textit{favorites}, and \textit{mentions}, in the aforementioned order, represent the user's popularity, content value, the preference among followers, and the user's name value. These metrics are collectively called influencing measures\cite{cha2010measuring} and is crucial in identifying the influence of a user over the network. Studies suggest that having millions of followers doesn't necessarily prove the influence of the user over the network and is known as \textit{A millions followers fallacy}\cite{cataldi201510}. Rather, an active audience who mentions a user, likes, and retweets his or her tweets, makes more contribution to the user's influence.

\textbf{Methodology for Comparing User Influence}

We evaluated the influence measure for each user and used the relative order of ranks as a measure of comparison for all 440,837 original users from the dataset. We sorted the users in decreasing order of their \textit{influence measure} where rank 1 indicated a user with the highest influence. The ranks assigned to measures were further used to analyze how ranks varied across various influencing measures and which categories of users were the top influencer for a measure.

We can utilise both the Pearson correlation coefficient or Spearman rank correlation coefficient to measure the strength of an association between two variables. Spearman rank correlation is preferred over Pearson correlation as it can capture the non-linear association between two variables while the latter can only capture the linear relation. Also, Pearson correlation works better on normally distributed data which is quite not the scenarios as visible in \textbf{Fig. }\ref{fig:scatterplots}. The Spearman does not require data to be normally distributed and is better suited to the need. Spearman correlation coefficient can be calculated by using Eq.\ref{eq:spearman} where $X_{i}$ and $Y_{i}$ are the ranks of users based on two different influence measures in a dataset of $N$ users. A perfect positive correlation of +1 or a negative correlation of -1 occurs when each of the variables is a perfect monotone function of the other.

\begin{equations}%
\rho = 1 - \frac{6\Sigma (X_{i} - Y_{i})^{2}}{N(N^{2}-1)}
\label{eq:spearman}
\end{equations} 

To investigate the correlation between four influencing measures for 4,40,837 original users, we calculated the Spearman rank correlation coefficient between each pair of measures for each regional language as well as for all languages combined as shown in  \textbf{Table }\ref{table:spearman}. A moderately high correlation (above 0.5) exists across the combinations of \textit{mention}, \textit{retweet}, and \textit{favourite} measures. It indicates that, in general, the users who are mentioned and liked more often are most retweeted. While there is a positive correlation of \textit{followers} with the other three measures, it doesn't appear to be as prominent as with a combination of the other three measures. This indicates that users with the most followers are not necessarily most mentioned. Also, they may not always produce content that is liked or retweeted most often. Effectively, the users with the most connections may not necessarily be the most influential people in terms of engaging the audience and having a significant outreach to the masses.

\begin{table*}[t]
\caption{Spearman's Rank-Order Correlation Coefficients}
	\label{table:spearman}
	\scriptsize
\centering
\resizebox{\linewidth}!{
\begin{tabular}{|l|l|l|l|l|l|l|l|l|l|l|l|l|l|}
\hline
\multicolumn{14}{|l|}{\textit{Language Codes: Hindi - hi, Tamil - ta, Urdu - ur, Marathi - mr, Telugu - te, Gujarati - gu}} \\ 
\multicolumn{14}{|l|}{\textit{Kannada - kn, Bengali - bn, Malayalam - ml, Oriya - or, Punjabi - pa, Sindhi - sd, All Languages - all}} \\ \hline
\diagbox{\textbf{Correlation}}{\textbf{Language}}           & \textbf{hi} & \textbf{ta} & \textbf{ur}  & \textbf{mr} & \textbf{te} & \textbf{gu} & \textbf{kn} & \textbf{bn} & \textbf{ ml} & \textbf{or} & \textbf{pa} & \textbf{sd} & \textbf{all} \\ \hline
\textbf{Follow-Favorite}  & 0.388 & 0.492 & 0.471 & 0.453   & 0.587  & 0.427    & 0.58    & 0.51    & 0.561     & 0.579 & 0.395   & 0.501  & 0.41          \\ \hline
\textbf{Follow-Retweet}   & 0.37  & 0.476 & 0.447 & 0.427   & 0.539  & 0.399    & 0.502   & 0.355   & 0.282     & 0.48  & 0.384   & 0.449  & 0.393         \\ \hline
\textbf{Follow-Mention}   & 0.462 & 0.524 & 0.492 & 0.498   & 0.586  & 0.475    & 0.563   & 0.445   & 0.384     & 0.591 & 0.458   & 0.502  & 0.477         \\ \hline
\textbf{Favorite-Retweet} & 0.641 & 0.769 & 0.696 & 0.67    & 0.745  & 0.677    & 0.756   & 0.578   & 0.601     & 0.664 & 0.647   & 0.707  & 0.657         \\ \hline
\textbf{Favorite-Mention} & 0.591 & 0.711 & 0.649 & 0.648   & 0.712  & 0.657    & 0.734   & 0.573   & 0.614     & 0.672 & 0.607   & 0.697  & 0.61          \\ \hline
\textbf{Retweet-Mention}  & 0.782 & 0.843 & 0.834 & 0.807   & 0.855  & 0.82     & 0.852   & 0.778   & 0.807     & 0.744 & 0.782   & 0.84   & 0.791         \\ \hline
\end{tabular}}
\end{table*}

\begin{table}[t]
\caption{Influencers Category and Corresponding Frequency}
	\label{table:followers}
	\scriptsize
\centering
\resizebox{\linewidth}!{
\begin{tabular}{|l|p{0.3\linewidth}|l|p{0.3\linewidth}|}
\hline
\multicolumn{4}{|l|}{\textit{O - Overall Count, Fo - Followers, M - Mentions, R - Retweets, Fa - Favorite}} \\ \hline
\textbf{Categories}       & \textbf{O, Fo, M, R, Fa}      & \textbf{Categories}      & \textbf{O, Fo, M, R, Fa}      \\ \hline
Politics                  & 65,32,28,29,30      & Religious                & 5,0,4,1,1                    \\ \hline
Media Org.                & 54,18,40,2,1        & Sports                   & 4,2,0,2,0                    \\ \hline
Media Person              & 29,9,13,20,18       & NA                       & 4,0,1,2,1                    \\ \hline
Entertainment             & 26,8,4,15,17        & Others                   & 2,0,1,1,0                    \\ \hline
Actor                     & 20,12,0,12,15       & Advocate                 & 2,0,1,2,1                    \\ \hline
Cricket                   & 15,14,0,3,8         & YouTuber                 & 1,1,0,0,0                    \\ \hline
Law                       & 8,2,3,3,4           & Suspend                  & 1,0,0,1,0                    \\ \hline
Health                    & 6,0,4,2,1           & NGO                      & 1,0,1,0,0                    \\ \hline
Corporate                 & 6,2,0,4,2           & Activist                 & 1,0,0,1,1                    \\ \hline

\end{tabular}}
\end{table}

The influential users may fall into various categories of people or organizations based on their profession. We collected the Twitter profiles of the top 100 users of each measure to identify the category where the most influential users might belong. These users were categorized into high-level categories such as \textit{people}, or \textit{organizations} belonging to \textit{politics, media, entertainment, and sports, etc.} The data from \textbf{Table} \ref{table:followers} shows that most users across influencing categories belong to people or organizations related to politics (\textit{Narendra Modi, PMO India, Amit Shah, Rahul Gandhi, Arvind Kejriwal, etc.}), sports and entertainment industry such as comedian, musicians, actors, cricketers (\textit{Salman Khan, Virat Kohli, Kapil Sharma, Kumar Vishwas, Filmfare, Saloni Gaur, etc.}), media persons or organizations (\textit{NDTV, Times of India, Aaj Tak, ABP News, Rajat Sharma, Sweta Singh, Sudhir Chaudhary, etc.}). Most of the users in the top 100 across measures are verified.

Users with a large number of followers get a lot of public attention owing to the fact they are directly connected to people in large number. Those with a relatively higher number of retweets or favourites have more engaging content that people tend to like or even retweet to propagate information further. On the other hand, users mentioned are notably political dignitaries such as the prime minister and union ministers of the country, chief ministers of various states and media persons or organizations. This indicates a deeper level of engagement or communication among users. It could be in response to the various government policies enacted during the pandemic, people voicing their opinions on the latest policy updates or advisories issued by the government or health organizations, or seeking help from individuals or organizations in emergencies.

We extracted the top 20 frequently occurring mentions from the \textit{IRLCov19} as shown in \textbf{Table }\ref{table:topMentions}. The findings show that people who post content in their regional languages generally prefer to mention regional media channels, local or state leaders and authorities.  The analysis could be useful to identify local leaders and authorities that could eventually help raise awareness and propagate help to the masses during the pandemic. The data further shows that most of the users mentioned are political dignitaries, be it regional ministers,  chief ministers of states, prime minister of the country,  along with the media persons or organizations. A significant portion of the mentioned users is verified, while those retweeting or posting the tweets are mostly non-verified.

The top influential users, across all four influential measures, are mostly pre-eminent public figures. Further, the top 100 users in each category show a significant overlap with one or the other. A combined list of the top 100 users from each category contains just 250 unique users.  We exploited the inference drawn earlier about the three influential measures - \textit{mention}, \textit{retweet} and \textit{favorite} showing the highest correlation among them to pick up the top twenty mentions across various regional languages as shown in \textbf{Table }\ref{table:topMentions}. The data shows that local leaders dominate in their corresponding region as per the regional language spoken. We have categorized these mentions in various categories as indicated on the top of the table. Each user in the table belongs to the category as indicated by the symbol in () and a verified tag, used as subscript \textit{v}, corresponding to a verified Twitter profile. The prime ministers of countries and those related to the prime minister's office are shown in \textbf{\textit{bold italics}}. The chief ministers, deputy chief ministers and their offices are shown in \textbf{bold only}. For example - \textbf{\textit{narendramodi}} ({P}$_{v}$) indicates that the user \textit{narendramodi} is a PM, has a verified Twitter profile and is related to politics.

\begin{table*}[t]
\caption{Top 20 Mentions Corresponding To Indian Regional Languages (\textit{IRLCov19})}
	\label{table:topMentions}
	\scriptsize
\centering
\resizebox{\linewidth}!{
\begin{tabular}{|l|l|l|l|} \hline

\multicolumn{4}{|l|}{\textit{P - Related to Politics, MP/MO - Media Person/Org., HP/HO - Health Person/Org.}}  \\ 
\multicolumn{4}{|l|}{\textit{G/O - Govt. Org./Org., A - Artist, L - Related to Law, C - Corporate Person, R - Related to Religion}}  \\
\multicolumn{4}{|l|}{\textit{N - NGO, S - Account Suspended, W - Account Withheld, NA - Account Doesn't Exist, OT - Others}}  \\\hline \hline
\textbf{} &  \textbf{} &  \textbf{}  &  \textbf{}  \\
\textbf{HINDI} & \textbf{TAMIL}  & \textbf{URDU}  & \textbf{MARATHI} \\ \hline
\textbf{\textit{narendramodi}} ({P}$_{v}$)   & pttvonlinenews ({MO}$_{v}$) & siasatpk ({MO}$_{v}$)  & rajeshtope11 ({HP}$_{v}$)  \\\hline
aajtak ({MO}$_{v}$)   & news7tamil ({MO}$_{v}$) & dawn\_news ({MO}$_{v}$)  & mahadgipr ({G}$_{v}$)   \\\hline
zeenews ({MO}$_{v}$)  & thatstamil ({MO}$_{v}$) & urduvoa ({MO}$_{v}$)  & \textbf{cmomaharashtra} ({P}$_{v}$)  \\\hline
dchaurasia2312 ({MP}$_{v}$)  & \textbf{cmotamilnadu} ({P}$_{v}$)  & bolnetwork ({MO}$_{v}$)  & abpmajhatv ({MO}$_{v}$)   \\\hline
\textbf{\textit{pmoindia}} ({P}$_{v}$)      & polimernews ({MO}$_{v}$) & nabthedentist ({HP}) & pawarspeaks(  {P}$_{v}$)   \\\hline
abpnews ({MO}$_{v}$)     & mkstalin ({P}$_{v}$)   & arynewsud ({MO}$_{v}$)   & mataonline ({MO}$_{v}$)    \\\hline
1stindianews ({MO})   & sunnewstamil ({P}$_{v}$)   & maizahameed ({P})   & dev\_fadnavis ({P}$_{v}$)  \\\hline
lambaalka ({P}$_{v}$)    & news18tamilnadu ({MO}$_{v}$)  & dunyanews ({MO}$_{v}$)   & \textbf{officeofut} ({P}$_{v}$)    \\\hline
\textbf{chouhanshivraj} ({P}$_{v}$) & thanthitv ({MO}$_{v}$)   & sheikhsafina ({A})  & zee24taasnews ({MO}$_{v}$)     \\\hline
opindia\_in ({MO})  & ishatamil ({N})    & maleehahashmey ({MP}$_{v}$)  & loksattalive ({MO}$_{v}$)   \\\hline
\textbf{myogiadityanath} ({P}$_{v}$)  & kalaignarnews ({MO})  & \textbf{\textit{imrankhanpti}} ({P}$_{v}$) & supriya\_sule ({P}$_{v}$)   \\\hline
mohfw\_india ({HO}$_{v}$)   & tamilthehindu ({MO}$_{v}$)  & gnnhdofficial ({MO}) & marathi\_rash ({MP})   \\\hline
ndtvindia ({MO}$_{v}$)    & jayapluschannel ({MO}$_{v}$)  & ptiofficial ({P}$_{v}$)   & bbcnewsmarathi ({MO}$_{v}$)   \\\hline
drharshvardha ({HP}$_{v}$) & dinakaranonline ({MO}$_{v}$)  & dr\_firdouspti ({P}$_{v}$)  & dgpmaharashtra ({L}$_{v}$)   \\\hline
\textbf{ashokgehlot51} ({P}$_{v}$)   & drramadoss ({P}$_{v}$)  & fawadchaudhry ({P}$_{v}$) & \textbf{\textit{narendramodi}} ({P}$_{v}$)    \\\hline
\textbf{arvindkejriwal} ({P}$_{v}$) & rajinikanth ({A}$_{v}$)  & hamidmirpak ({MP}$_{v}$)   & anildeshmukhncp ({P}$_{v}$)  \\\hline
vikasbhaabp ({NA})  & vikatan ({MP}$_{v}$) & bbcurdu ({MP}$_{v}$)   & smartpune ({O}) \\\hline
drkumarvishwas ({A}$_{v}$) & aloor\_shanavas{P}$_{v}$) & psppakistan ({P}$_{v}$) & milokmat ({MO}$_{v}$)   \\\hline
ashutosh83b ({MP}$_{v}$)  & \textbf{\textit{narendramodi}} ({P}$_{v}$)     & \textbf{\textit{pakpmo}} ({P}$_{v}$)   & \textbf{ajitpawarspeaks} ({P}$_{v}$)   \\\hline
sardanarohit ({MP}$_{v}$)   & arjunsaravanan5 ({L})  & tahirulqadriur ({R})  & \textbf{\textit{pmoindia}} ({P}$_{v}$)  \\ \hline \hline
\textbf{} &  \textbf{} &  \textbf{}  &  \textbf{}   \\

\textbf{TELUGU}   & \textbf{GUJARATI} & \textbf{KANNADA} &  \textbf{BENGALI} \\ \hline
ntvjustin ({MO})  & vtvgujarati ({MO}) & \textbf{bsybjp} ({P}$_{v}$)     & {banglargorbomb} ({P}$_{v}$)  \\\hline
arogyaandhra ({HO}$_{v}$) & \textbf{vijayrupanibjp} ({P}$_{v}$) &\textbf{cmofkarnataka} ({P}$_{v}$)   & abpanandatv ({MO}$_{v}$)  \\ \hline
janasenaparty ({P}$_{v}$)  & news18guj ({MO}$_{v}$) & siddaramaiah ({P}$_{v}$)   & bbcbangla ({MO}$_{v}$)    \\ \hline

bbcnewstelugu ({MO}$_{v}$)  & \textbf{\textit{narendramodi}}  ({P}$_{v}$) & srisamsthana ({R})   & bjp4bengal ({P}$_{v}$) \\ \hline
\textbf{ysjagan} ({P}$_{v}$)  & \textbf{cmoguj} ({P}$_{v}$)   & sriramulubjp ({P}$_{v}$) & ei\_samay ({MO}$_{v}$)   \\ \hline
pawankalyan ({P}$_{v}$)   & sandeshnews ({MO})  & publictvnews ({MO}$_{v}$)  & didikebolo ({P}$_{v}$)   \\ \hline
\textbf{telanganacmo} ({P}$_{v}$)  & ddnewsgujarati ({MO}$_{v}$) & suvarn ({OT})  & \textbf{mamataofficial} ({P}$_{v}$) \\ \hline
tarak9999 ({A}$_{v}$)   & divya\_bhaskar ({MO}$_{v}$) & kumarishobakka ({S}) & dailystarnews ({MO}$_{v}$)  \\ \hline
 jaitdp ({P}$_{v}$)     & bsnl\_gj ({G}$_{v}$)   & prajavani ({MO}$_{v}$)    & airnews\_ghy ({MO}$_{v}$)   \\ \hline
urstrulymahesh ({A}$_{v}$) & jitu\_vaghani ({P}$_{v}$)  & \textbf{\textit{narendramodi}} ({P}$_{v}$)     & cpkolkata ({L}$_{v}$)    \\ \hline
 \textbf{\textit{narendramodi}} ({P}$_{v}$)     & zee24kalak ({MO}$_{v}$)   & kicchasudeep ({A}$_{v}$)  & cpim\_westbengal ({P}$_{v}$) \\ \hline
 jspveeramahila ({P}) & tv9gujarati ({MO}$_{v}$)   & shakunthalahs ({P})  & aitcofficial ({P}$_{v}$)   \\ \hline
bharatysrcp ({P}$_{v}$)  & pibahmedabad ({MO}$_{v}$)   & bjp4karnataka ({P}$_{v}$)  & dw\_bengali ({MO}$_{v}$)  \\ \hline
 tv9telugu ({MO}$_{v}$)    & gujaratpolice ({L}$_{v}$)  & oneindiakannada ({MO}$_{v}$)  & news18bengali ({MO}$_{v}$) \\ \hline
bolisetti\_satya ({P}) & \textbf{nitinbhai\_patel} ({P}$_{v}$) & vijaykarnataka ({MO}$_{v}$) & \textbf{\textit{narendramodi}} ({P}$_{v}$) \\ \hline

jspshatagniteam ({MP})  & aravindchaudhri ({MP}$_{v}$)  & news18kannada ({MO}$_{v}$)  & kpeastsubndiv ({L}$_{v}$)  \\\hline
uttarandhranow ({MO}) & bjp4gujarat ({P}$_{v}$)  & inckarnataka ({P}$_{v}$)   & pib\_india ({MO}$_{v}$) \\\hline

 ncbn(P$_{v}$)     & sanjayezhava ({MP})   & anilkumble1074 ({A}$_{v}$) & myanandabazar ({MO}$_{v}$) \\\hline
 manvidad ({OT})   & gujratsamachar ({MO}$_{v}$) & puneethrajkumar ({A}$_{v}$)  & dailyittefaq ({MO})  \\\hline
 ktrtrs ({P}$_{v}$)     & collectorbk ({L})  & bbmpcomm ({L}$_{v}$)    & mohfw\_india ({HO}$_{v}$)  \\\hline
\textbf{} &  \textbf{} &  \textbf{}  &  \textbf{} \\

\textbf{ORIYA}  &  \textbf{MALAYALAM} &  \textbf{PUNJABI} &  \textbf{SINDHI}  \\ \hline
kanak\_news ({MO}$_{v}$)   & \textbf{cmokerala} ({P}$_{v}$)   & jagbanionline ({MO}$_{v}$)  & khalidkoree ({OT}) \\\hline
news18odia ({MO})  & asianetnewstv ({NA})  & dailyajitnews ({MO})  & muradalishahppp ({P})  \\\hline
sambad\_odisha ({MO})  & \textbf{vijayanpinarayi} ({P}$_{v}$)   & \textbf{capt\_amarinder} ({P}$_{v}$)  & sialrabail ({S})   \\\hline
otvkhabar ({MO})   & manukumarjain ({C}$_{v}$)   & ptc\_network ({MO}$_{v}$) & mukhtar\_soomro ({OT})  \\\hline
 \textbf{naveen\_odisha} ({P}$_{v}$)  & pibtvpm ({MO}$_{v}$)    & ptcnews ({MO})    & mahamsindhi ({MP})  \\\hline
 \textbf{cmo\_odisha} ({P}$_{v}$)   & thekeralapolice ({L}$_{v}$)  & ishehnaaz\_gill ({A}$_{v}$)  & faraz\_aligg ({OT})   \\\hline
 hfwodisha ({HO}$_{v}$)   & manoramaonline ({MO}$_{v}$)   & punjabpmc ({P}$_{v}$)    & bbhuttozardari ({P}$_{v}$) \\\hline
 odishareporter ({MO}$_{v}$)  & nishthvanth ({OT})    & punjabgovtindia ({G}$_{v}$)  & ayazlatifpalijo ({P}$_{v}$)  \\\hline
 \textbf{\textit{narendramodi}} ({P}$_{v}$)    & news18kerala ({MO}$_{v}$)   & pib\_india ({MO}$_{v}$)   & dadu\_plus ({S})   \\\hline
ipr\_odisha ({MO}$_{v}$)   & sathyamaanu ({OT}) & mib\_india ({MO}$_{v}$)   & drhamadwassan ({OT}) \\\hline
drgynaec ({OT})    & thatsmalayalam ({MO}$_{v}$)   & gurmeetramrahim ({W})  & ahtishamqhala ({OT})  \\\hline
anandstdas ({MP})  & ddnewsmalayalam ({MO}$_{v}$)  & \textbf{\textit{narendramodi}} ({P}$_{v}$)     & abbasimehwish ({OT})  \\\hline
dpradhanbjp ({P}$_{v}$)   & zhmalayalam ({MO}) & newscheckerin ({MO}$_{v}$)  & sindhcmhouse ({P})   \\\hline
nandighoshatv ({MO}) & mfwaikerala ({OT})   & \textbf{cmopb} ({P}$_{v}$)     & sindhikhabroon ({MO}) \\\hline
zeeodisha ({MO})   & vikramanmuthu ({OT})  & pibchandigarh ({MO}$_{v}$)  & najeebabro2 ({P})  \\\hline
\textbf{\textit{pmoindia}} ({P}$_{v}$)      & avs\_ind ({OT})     & incpunjab ({P}$_{v}$)    & sangrisaeed ({OT}) \\\hline

skilledinodisha ({C}) & comrademallu ({P})   & harsimratbadal\_ ({P}$_{V}$) & sama4newz ({NA})    \\\hline
bjd\_odisha ({P}$_{v}$)   & ambath ({OT})     & punjabpoliceind ({L}$_{v}$)  & mnavax ({OT})     \\\hline

satyaparida01 ({OT}) & kavyasree19941 ({NA})   & sportsperson5 ({OT})  & chandio\_gs ({MP})  \\\hline
theargus\_in ({NA})  & manoramanews ({MO}$_{v}$)   & derasachasauda ({W}) & sindhicongress ({N})  \\ \hline

\end{tabular}}
\end{table*}

\section{DATASET ACCESS}\label{sec:dataset}
The dataset is accessible on GitHub\cite{Covid19IRLTDataset}. However, to comply with Twitter's content redistribution policy\cite{twitterDeveloperPolicy}, we are distributing only the IDs of the collected tweets. Tools such as Hydrator\cite{hydrator} can be used to retrieve the full tweet object.

 \clearpage
 
\section{CONCLUSION AND FUTURE WORK}\label{sec:conclusion} This paper presents \textit{IRLCov19} - a Twitter dataset of Indian regional languages on the Covid-19 pandemic. The dataset has been collected over a period of 6 months between Feb 01, 2020, to July 31, 2020, and consists of over 13 million multilingual tweets.
The tweets in the \textit{IRLCov19} are from more than 1.4 million Twitter users that includes more than 5k
verified users. The tweets in the dataset span 12 different Indian regional languages. The dataset can be advantageous for researchers, Government authorities, and policymakers in studying the pandemic from a varied perspective such as understanding public reactions and opinions, early detection and surveillance of the pandemic etc.

Identifying influencers and their locations is a crucial task amid a crisis or emergency. It paves the way for disease hotspot detection, employing suitable and effective information publishing strategies, and tracking and debunking misinformation floating in the network. We utilized GeoPy, a python library, to extract the location of a tweet and use it to collect relevant tweets from the identified location. We further exploited the collected information to identify the top local leaders or influencers and the profiles to which influencers belong. We plan to update the dataset with more paradigms about the COVID-19 dataset related to Indian Regional Languages. Future studies could explore the information-sharing behaviour among the users and how different groups respond to the pandemic.

\bibliographystyle{unsrt}
\bibliography{ms}

\end{document}